\documentclass{hep99}
\begin{document}

\title{A glueball-$q \overline q$ filter in central production}

\author{A. Kirk}
\address{School of Physics, Birmingham University,
         Birmingham, U.K.\\[3pt]
E-mail: {\tt andrew.kirk@cern.ch}}

\abstract{
A study of central meson production as a function of the
difference in transverse momentum ($dP_T$)
of the exchanged particles
shows that undisputed $q \overline q$ mesons
are suppressed at small $dP_T$ whereas the glueball candidates
are enhanced and that
the production cross section for different
resonances depends strongly on the azimuthal angle between the
two outgoing protons.
}

\maketitle


\section{Introduction}
\par
There is considerable current interest in trying to isolate the lightest
glueball.
Several experiments have been performed using glue-rich
production mechanisms.
One such mechanism is Double Pomeron Exchange (DPE) where the Pomeron
is thought to be a multi-gluonic object.
Consequently it has been
anticipated that production of
glueballs may be especially favoured in this process~\cite{closerev}.
\par
The WA102 experiment at the CERN Omega Spectrometer
studies centrally produced exclusive final states
formed in the reaction
\noindent
\begin{equation}
pp \longrightarrow p_{f} X^{0} p_s,
\label{eq:1}
\end{equation}
where the subscripts $f$ and $s$ refer to the fastest and slowest
particles in the laboratory frame respectively and $X^0$ represents
the central system.

\section{A glueball-$q \overline q$ filter in central production~?}
The WA102 experiment studies mesons produced in double exchange processes.
However, even in the case of pure DPE
the exchanged particles still have to couple to a final state meson.
The coupling of the two exchanged particles can either be by gluon exchange
or quark exchange. Assuming the Pomeron
is a colour singlet gluonic system if
a gluon is exchanged then a gluonic state is produced, whereas if a
quark is exchanged then a $q \overline q $ state is produced~\cite{closeak}.
In order to describe the data in terms of a physical model,
Close and Kirk~\cite{closeak},
have proposed that the data be analysed
in terms of the difference in transverse momentum ($dP_T$)
between the particles exchanged from the
fast and slow vertices.
The idea being that
for small differences in transverse momentum between the two
exchanged particles
an enhancement in the production of glueballs
relative to $q \overline q$ states may occur.
\begin{figure*}
 \vspace{10.0cm}
\begin{center}
\includegraphics{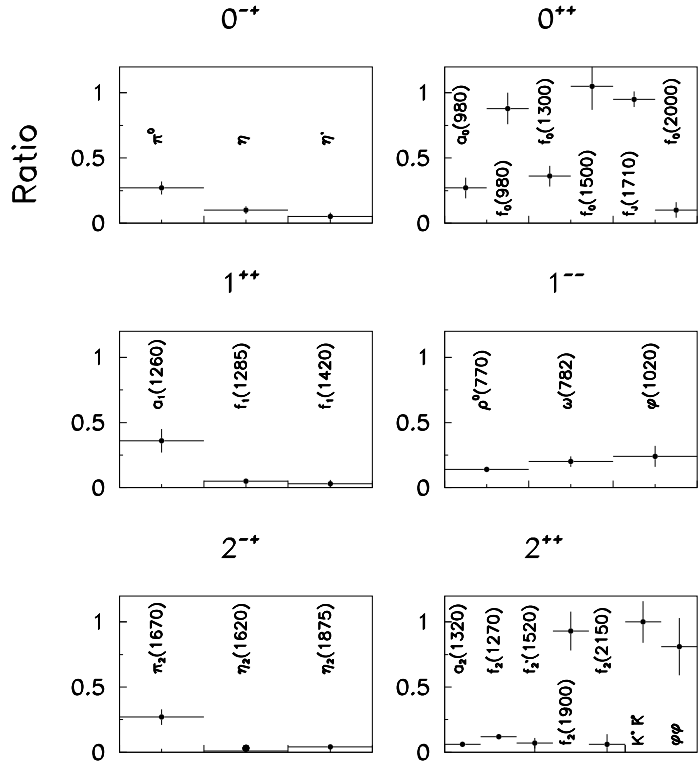}
\end{center}
\caption{ The ratio of the amount of resonance with
$dP_T$~$\leq$~0.2 to the amount with
$dP_T$~$\geq$~0.5~GeV.
}
\label{fracratio}
\end{figure*}
\par
The ratio of the number of events
for $dP_T$ $<$ 0.2 GeV to
the number of events
for $dP_T$ $>$ 0.5 GeV for each resonance considered has been
calculated~\cite{memoriam} and the
results are shown in fig.~\ref{fracratio}.
As can be
observed all the undisputed $q \overline q$ states
which can be produced in DPE, namely those with positive G parity and $I=0$,
have a very small value for this ratio ($\leq 0.1$).
Some of the states with $I=1$ or G parity negative,
which can not be produced by DPE,
have a slightly higher value ($\approx 0.25$).
However, all of these states are suppressed relative to the
the glueball candidates the
$f_0(1500)$, $f_J(1710)$, and $f_2(1900)$,
together with the enigmatic $f_0(980)$,
which have
a large value for this ratio~\cite{memoriam}.
\begin{figure*}
 \vspace{9.0cm}
\begin{center}
\includegraphics{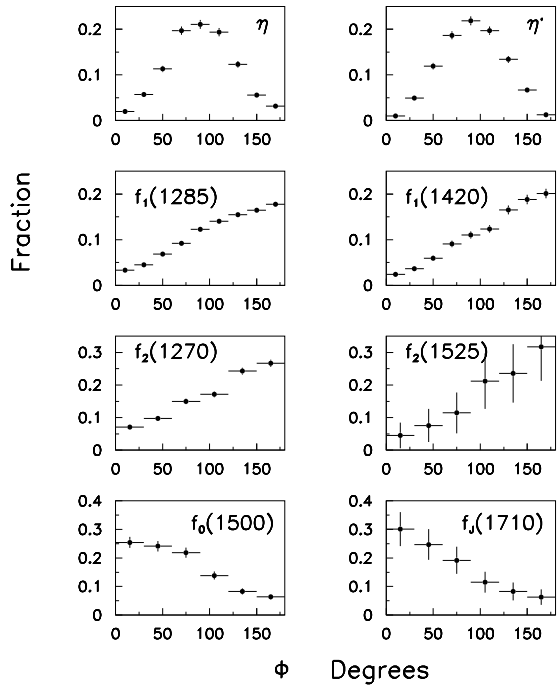}
\end{center}
\caption{The azimuthal angle between the fast and
slow protons $\phi$ for various resonances.
}
\label{fi:phidep}
\end{figure*}

\section{The azimuthal angle between the outgoing protons}

\par
The azimuthal angle $\phi$ is defined as the angle between the $p_T$
vectors of the two protons.
Naively it may be expected that this angle would be flat irrespective
of the resonances produced.
Fig.~\ref{fi:phidep}  shows the $\phi$ dependence for two resonances with
$J^{PC}$~=~$0^{-+}$ (the $\eta$ and $\eta^\prime$),
two with $J^{PC}$~=~$1^{++}$ (the $f_1(1285)$ and $f_1(1420)$),
two with $J^{PC}$~=~$2^{++}$ (the $f_2(1270)$ and $f_2^\prime(1525)$)
and two with $J^{PC}$~=~$0^{++}$ (the $f_0(1500)$ and $f_J(1710)$).
The $\phi$ dependence is clearly not flat and considerable variation
is observed between resonances with different $J^{PC}$s.
\begin{figure*}
 \vspace*{5.0cm}
\begin{center}
\includegraphics{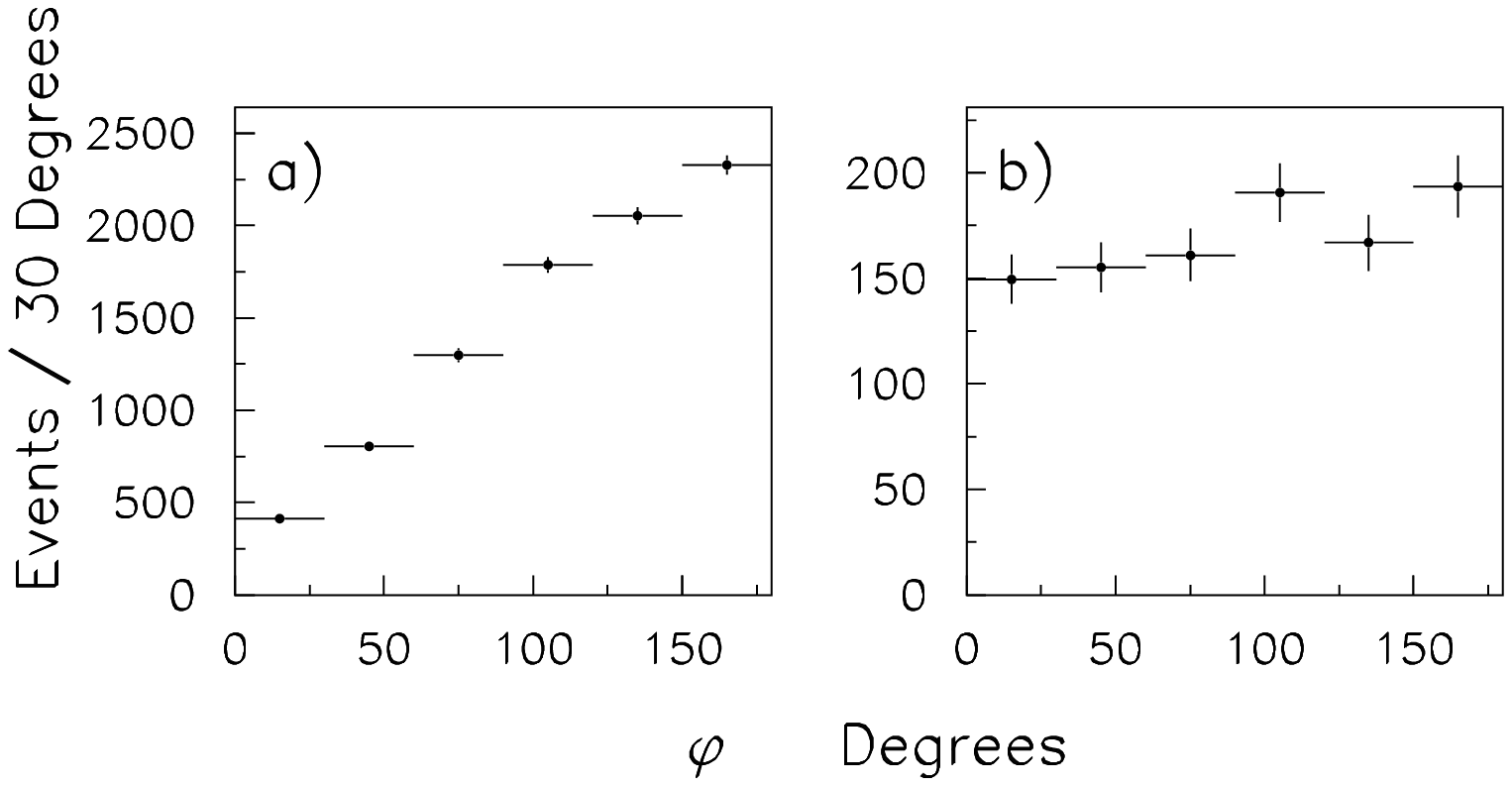}
\end{center}
\caption{The azimuthal angle between the fast and
slow protons  $\phi$  for the $f_1(1285)$ for a) $| t_1 - t_2 |$ $\le$ 0.2
and b) $|t_1 - t_2|$ $\ge$ 0.4
}
\label{fi:tdep}
\end{figure*}
\par
Several theoretical papers have been published on these
effects~\cite{angdist,clschul}.
All agree that the exchanged particle
(Pomeron) must have J~$>$~0
and that J~=~1
is the simplest explanation.
Using $\gamma^* \gamma^*$ collisions as an analogy
Close and Schuler\cite{clschul} have calculated the $\phi$ dependencies
for the production of resonances with different $J^{PC}$s.
They have found that for a $J^{PC}$~=~$0^{-+}$ state
\begin{equation}
\frac{d^3\sigma}{d\phi dt_1dt_2} \propto t_1 t_2 sin^2 \phi
\end{equation}
as can be seen from fig.~\ref{fi:phidep}
the $\phi$ distributions are proportional to $sin^2 \phi $ and
it has been found
experimentally
that $d\sigma/dt$ is proportional
to $t$ ~\cite{0mppap}.
For the  $J^{PC}$~=~$1^{++}$ states this model predicts that
$J_Z$~=~$\pm1$ should dominate, which has been found to be
correct~\cite{f1pap}, and
\begin{equation}
\frac{d^3\sigma}{d\phi dt_1dt_2}\propto (\sqrt{t_2} - \sqrt{t_1})^2 + \sqrt{t_1
t_2}sin^2 \phi/2
\label{1pp}
\end{equation}
As can be seen from fig.~\ref{fi:phidep}
the $\phi$ distributions are proportional to
$\alpha + \beta sin^2 \phi/2 $. In addition
equation(\ref{1pp})
would predict that when $|t_2 - t_1|$ is small
$d\sigma/d\phi$ should be proportional to $sin^2 \phi/2 $ while
when
$|t_2 - t_1|$ is large
$d\sigma/d\phi$ should be constant. As shown in fig.~\ref{fi:tdep} this
trend is observed in the data.
The aim now is to study the $\phi$ dependences of other
known $q \overline q$ states in order to understand
more about the nature of the Pomeron and then to use this
information as a probe for non-$q \overline q$ states.
\section{Summary}
\par
A study of centrally produced pp interactions
show that there is the possibility of a
glueball-$q \overline q$ filter mechanism ($dP_T$).
All the
undisputed $q \overline q $ states are observed to be suppressed
at small $dP_T$, but the glueball candidates
$f_0(1500)$, $f_J(1710)$, and $f_2(1900)$ ,
together with the enigmatic $f_0(980)$,
survive.
In addition, the production cross section for different
resonances depends strongly on the azimuthal angle between the
two outgoing protons which may give information on the nature
of the Pomeron.

\end{document}